\documentclass[a4paper,12pt,titlepage]{article}

\usepackage[affil-it]{authblk}
\usepackage{graphicx}
\usepackage{subfigure}
\usepackage{psfrag}
\usepackage{hyperref}
\usepackage[super,nonamebreak]{natbib}
\usepackage{amsmath}
\usepackage{amssymb}
\usepackage{setspace}
\usepackage{chemarr}
\begin{document}

\title{\textbf{ NEURONAL SPACE CONSTANTS IN VARIOUS GEOMETRIES AND PHYSIOLOGICAL CONDITIONS}}
%USING \TeX\ OR \LaTeX\footnote{For the title, try not to use more than
%3 lines. Typeset the title in 11~pt bold and uppercase.}

\author {ASHA GOPINATHAN }
\affil{ Department of Neurology,Sree Chitra Tirunal Institute for Medical Sciences and Technology, Trivandrum $695011$, Kerala, India}
\affil{Tel: +919633568106)}
\affil{dendron.15@gmail.com}
\date { }
%Typeset names in 9~pt roman, uppercase. Use the footnote to indicate the
%present or permanent address of the author.}}

%\address{University Department, University Name, Address\\
%City, State ZIP/Zone,Country\,\footnote{State completely without
%abbreviations, the affiliation and mailing address,
%including country. Typeset in 9~pt italic.}\\
%\email{author\_id@domain\_name\footnote{Typeset author e-mail address
%in single line.}}
%\http{$<$webaddress$>$}
%}

\maketitle

%\begin{history}
%\received{Day Month Year}
%\revised{Day Month Year}
%\end{history}
\section*{\textbf {Abstract}}
\begin{normalsize}
The cable equation is a second order, parabolic, partial differential equation that describes the evolution of voltage in the dendrite of a neuron. Here we look at the various ways in which lambda ( space constant/variable space constant/ length parameter ) is calculated in the cases of linear, passive transmission as well as nonlinear active transmission. Changes in morphology are taken into account by including the case of tapering dendrite, branched dendrites, branched dendrites with taper or flare. The case of variable membrane resistance and the relationship between input resistance and space constant is explored. Finally the reaction diffusion equation governing the diffusion of calcium in dendrites and space constant associated with that is described.   \\
%the abstract in 9~pt roman with baselineskip of 11~pt, making
%an indentation of\break
%0.25 inches on the left and right margins.
\textit{Keywords}: cable equation; reaction diffusion equation; space constant; variable space constant; length parameter
\end{normalsize}
\vspace{2mm}
\newline
\section{\textbf {Introduction}}
\begin{large}
When a cell is isopotential, the membrane potential is uniform at all points of the cell, depending only on time. This is apt for describing signalling in a cell body which can be assumed to be a sphere [8,21].  However if one needs to look at electrical or diffusional properties of  dendrites or axons, one needs to approximate these by a cylinder ( or more if branched). Here the geometry of the cell plays a role. In this context, space plays a part along with time. Thus a factor called space constant is defined which is related to the diameter $d$ of the cell and thus its geometry. In this paper, the notion of space constant is explored under various conditions : from passive to active, unbranched to branched and nontapered to tapered processes. Space constant can also be variable due to changes in membrane resistance $R_{m}$.\\
In the case of a passive infinite and semi infinite cable, space constant ( $\lambda$) is the distance where voltage reaches $37\%$ ($\frac{1}{e}$ )of its original.  When electrotonic length $L=1$, the voltage will decay to only $65 \%$ of its original value at $x = \lambda$. This is due to the boundary conditions at $x=\lambda$. In the infinite cable an infinite cylinder with its associated conductance is attached to the artificial boundary at $ x= \lambda$ and current flows into this cylinder, whereas in a finite cylinder with $L=1$ the cylinder ends at $x=\lambda$. For a sealed end, there is no current flow and conductance at this boundary. Voltage decay is also affected by taper or flare in the main cable or its branches [4]. When $\lambda$ is large, the spatial decay of input with distance is small and vice versa for small $\lambda$.  On the other hand in the case of active or sinusoidal stimulation, $\lambda_{\omega}$ has the dimension of interelectrode distance when the peak value of sinusoidal transmembrane distance is attenuated to $\frac{1}{e}$.\\
It is seen that $\lambda $ for both semi infinite and finite cable is directly proportional to input resistance. Furthermore $\lambda $ in an active cell or with sinusoidal current injection is proportional to $\sqrt Z$ ( impedance) and $\sqrt \frac {1}{f}$ (frequency).   
\end{large} 
\vspace{1mm}
\section{\textbf{Unbranched dendrite}}
\begin{flushleft}
\begin{large}
Linear cable theory assumes that the cable has a uniform diameter and leakage resistance. So it can be used in the passive case where there is no time or voltage dependent conductances. When $R_{m}$, the specific membrane resistance is much larger than $R_{i}$,the intracellular resistivity, it can be safely assumed that all current flow will be one-dimensional along the length of the cable. Under such conditions there are equations describing the change in voltage across space and time along the cable.
\end{large}
\end{flushleft}
\vspace{2mm}
\begin{flushleft}
\begin{large}
 The cable equation is :
\begin{equation}
\frac{1}{r_{a}}\frac{\partial^{2}V_{m}(x,t)}{\partial x^{2}}= \frac{V_{m}(x,t)- V_{rest}}{r_{m}} + c_{m}\frac{\partial V_{m}}{\partial t} - I_{inj}(x,t)
\end{equation}
where $r_{a}$ is the intracellular resistance per unit length of cable with dimensions of ohms per centimeter and $r_{m}$ is the membrane resistance of a unit length of fiber measured in units of ohms-centimeter.
Rewriting this, 
\begin{equation}
 \frac{r_{m}}{r_{a}}\frac{\partial^{2}V_{m}(x,t)}{\partial x^{2}} = V_{m}(x,t)- V_{rest}+ r_{m}c_{m}\frac{\partial V_{m}}{\partial t} - r_{m}I_{inj}(x,t)
\end{equation}
This leads to ,
\begin{equation}
 \lambda^{2}\frac{\partial^{2}V_{m}(x,t)}{\partial x^{2}} = \tau \frac{\partial V_{m}}{\partial t} + V_{m}(x,t)- V_{rest}- r_{m}I_{inj}(x,t)
\end{equation}
\end{large}
\end{flushleft}

\subsection{\textbf\it{$\lambda_{DC}$:}}
\begin{large}
\begin{equation}
 \lambda_{DC} = (\frac{r_{m}}{r_{a}})^{\frac{1}{2}}
\end{equation}
As $ r_{m}= \frac{R_{m}}{\pi d} $ and $ r_{a} = \frac{4R_{i}}{\pi d^2} $
\begin{equation}
\lambda_{DC}= (\frac{R_{m}d}{R_{i}4})^{1/2}
\end{equation}
where $d$ is the diameter of the cable or dendrite as the dendrite is modeled as a cable. 
\begin{equation}
 V(x) = V_{0}exp{(\frac{-x}{\lambda_{DC}})}
\end{equation} 
where $V_{0}$ is a constant or very slowly varying voltage applied to one end of a long cable. See Figure 1. 
\end{large}
\begin{figure}[!ht]
\begin{center}
\psfrag{Stim}{$Stim$}
\psfrag{R}{$R$}
\psfrag{lambda}{$lambda$}
\subfigure[{dendrite}]{
\includegraphics[width = 5.0in]{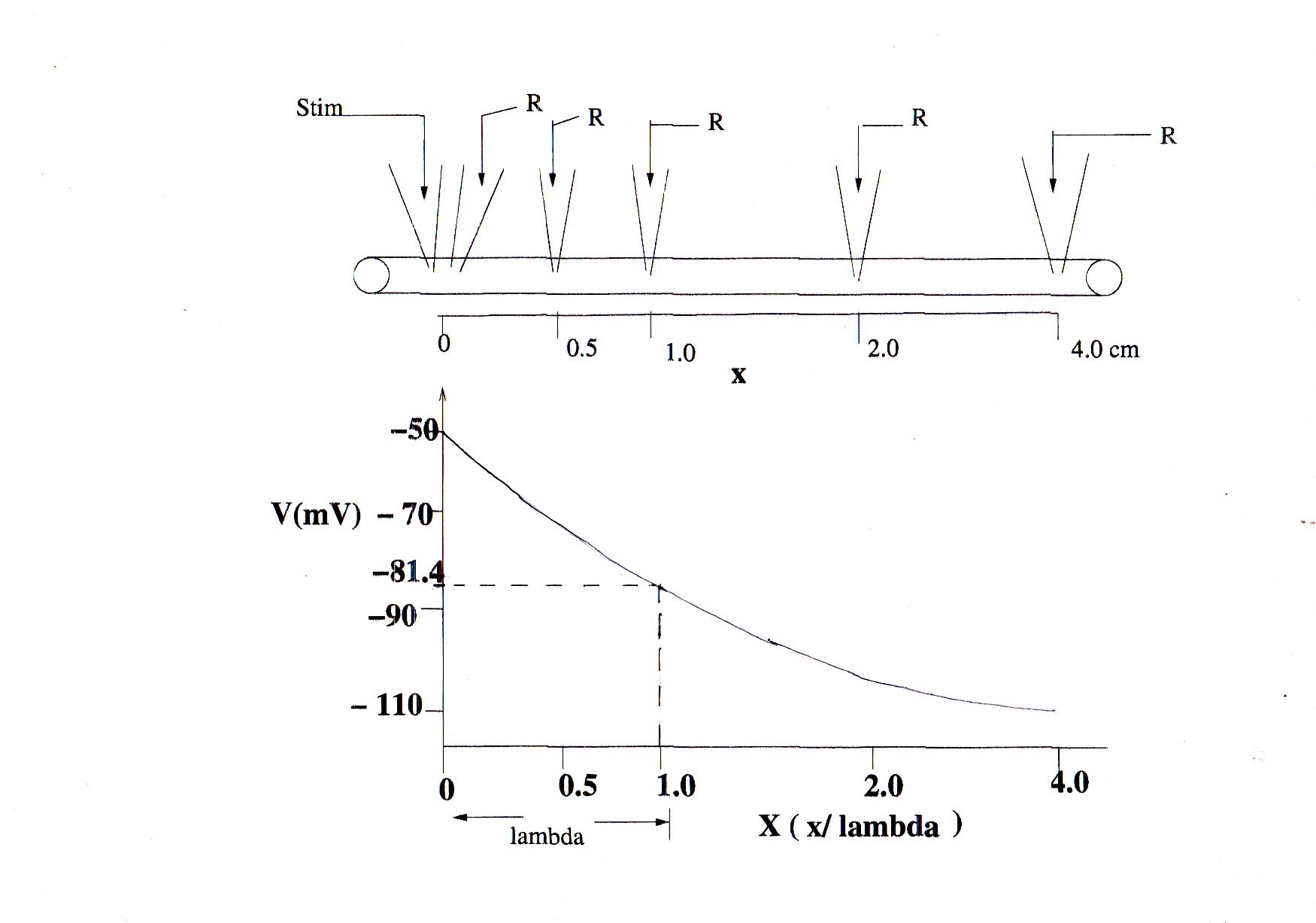}
}
\psfrag{Stim}{$Stim$}
\psfrag{R}{$R$}
\psfrag{lambda}{$lambda$}
\psfrag{Sealed end}{$Sealed end$}
\psfrag{L=x/lambda=1.0}{$L=x/lambda =1.0$}
\subfigure[{dendrite with sealed end}]{
\includegraphics[width =5.0in]{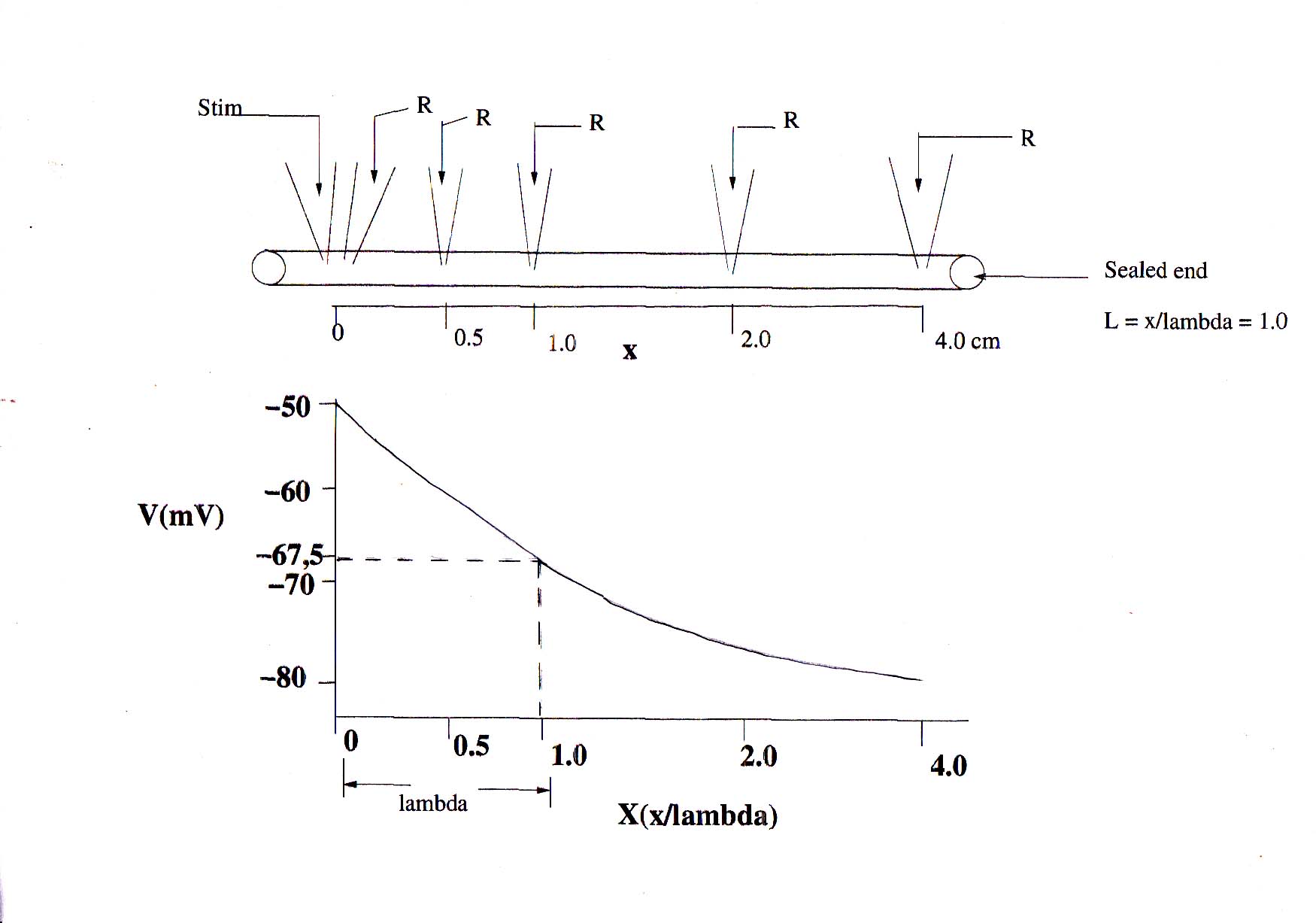}
}
\end{center}
\caption{\textbf{Comparison of cable with and without boundary condition }}
\label{fig:lambda}
\end{figure}
\vspace{2mm}
\subsection{\textbf\it{$\lambda_{DCtaper}$:}}
\begin{large}
 In a tapering dendrite $d$ changes and thus the $\lambda_{DCtaper}$ should reflect that. 
From  [14,equation 3 ]
\begin{equation}
 \lambda_{DCtaper}= \lambda_{DC}\biggl[\frac{r}{r_{0}}\biggr]^{1/2}\biggl[1+ (\frac{dr}{dx})^{2}\biggr]^{-1/4}
\end{equation} 
where $r$ is the radius of the cylinder at $x$, $x$ is the distance from the soma, $r_{0}$ is the radius of the cylinder at $ x=0 $.
This can be derived as given below.  For a dendrite with taper the cable equation is [5 equation 3] : 
\begin{equation}
 V + \tau \frac{\partial V}{\partial t} = \frac{R_{m}r}{2R_{i}}(\frac{\partial^{2}V}{\partial x^{2}} + \frac{\partial V}{\partial x}\frac{2}{r}\frac{dr}{dx})(\frac{ds}{dx})^{-1}
\end{equation} 
\begin{equation}
 \frac{ds}{dx} = ( 1 + (\frac{dr}{dx})^{\frac{1}{2}})^{\frac{1}{2}}
\end{equation} 
Substituting this in equation $8$ and multiplying and dividing by $r_{o}$ gives 
\begin{equation}
 V + \tau \frac{\partial V}{\partial t} = \frac{R_{m}r_{o}}{2R_{i}} \frac{r}{r_{o}}(\frac{\partial^{2}V}{\partial x^{2}} + \frac{\partial V}{\partial x}\frac{2}{r}\frac{dr}{dx})(1+ (\frac{dr}{dx})^{2})^(\frac{-1}{2})
\end{equation} 
This can be written as :
\begin{equation}
 V + \tau \frac{\partial V}{\partial t} = (\lambda_{DC})^{2}\frac{r}{r_{o}}(1+ (\frac{dr}{dx})^{2})^(\frac{-1}{2})(\frac{\partial^{2}V}{\partial x^{2}}+\frac{\partial V}{\partial x}\frac{2}{r}\frac{dr}{dx})
\end{equation} 
From here it can be deduced that :
\begin{equation}
 (\lambda_{tap})^{2} = (\lambda_{DC})^{2}\frac{r}{r_{o}}(1+ (\frac{dr}{dx})^{2})^(\frac{-1}{2})
\end{equation} 
Thus, 
\begin{equation}
 \lambda_{tap} = (\lambda_{DC})(\frac{r}{r_{o}})^{\frac{1}{2}}(1+ (\frac{dr}{dx})^{2})^(\frac{-1}{4})
\end{equation} 
If the tapering dendrite is sufficiently long to be considered semi infinite ($0\leq x < \infty$) or infinite ($-\infty < x <\infty$), the solutions can be obtained in closed form. The boundary conditions for the semiinfinite case can be any of the following :  voltage clamp, sealed end, killed end, current injection at one end, natural termination, lumped-soma termination. For infinitely large x , V remains bounded. Thus $ lim_{|x|\to \infty} |V(x,t)|< \infty, t>0 $[28]. 
\end{large} 
\vspace{2mm}
\newline
\subsection{\textbf\it{$\lambda_{DCvariable}$:}}
\begin{large}
 Under certain conditions,$R_{m}$ is not uniform but variable across the dendrite [11]. Here the cable equation becomes :
\begin{equation}
 (\lambda)^{2}(x)\frac{\partial^{2}V}{\partial x^{2}} - \tau_{m}(x)\frac{\partial V}{\partial t} - V = 0
\end{equation} 
where\begin{equation}
   \lambda_{DC}(x) = \sqrt{\frac{R_{m}(x)d}{4 R_{i}}}  
     \end{equation}   is the variable space constant. 
When $R_{m}(x)$ = constant, $\lambda_{DC}(x) = \lambda_{DC}$ \\
The electrotonic length of a dendrite is a measure of its length in nondimensional terms. It is written as $ L = \frac{l}{\lambda} $ where $l$  is the length.\\
London et al [11] show that if the total membrane conductance( $G_{m}=1/R_{m}$) is kept fixed for a given cable,then any nonuniformity of $G_{m}$ over the cylinder reduces the electrotonic length L. If it is a sealed end cable, any monotonic increase in $G_{m}$ improves the voltage transfer from input location to the soma. It is further improved if it is synaptic input at a distal location as opposed to a current source.\\  
The input resistance of an infinite cable is the following ratio in the limit that the distance between the current passing electrode and recording electrode shrinks to zero [9 equations$ 2.14,2.15 $]
 \begin{equation}
  R_{in} = \frac{V(x)}{I_{i}(x)} = \frac{V(x=0)}{I_{0}}                                            
\end{equation} 
 This can be written as :
\begin{equation}
 R_{in}= \frac{r_{m}}{2\lambda} = \frac{r_{a}\lambda}{2} = \frac{(r_{a}r_{m})^{\frac{1}{2}}}{2}
\end{equation} 
For a semi- infinite cable, $R_{\infty}$ is double that of an infinite cable as there are two semi-infinite cables to one infinite cable [9, equation $2.16$]
\begin{equation}
 R_{\infty}= (r_{a}.r_{m})^{\frac{1}{2}} = r_{a}\lambda = \frac{r_{m}}{\lambda} = \frac{((R_{m}R_{i})^{\frac{1}{2}})2}{\pi d^{\frac{3}{2}}}
\end{equation} 
In multidimensional systems like dendritic trees, cardiac, smooth and skeletal muscle, current spread is in more than one dimension . Here an effective 'input resistance' is defined as a function of electrode separation $r$. In the steady state for a 2-D case [8 equations 5.17,5.18 ]
\begin{equation}
 \frac{V}{I_{o}} = \frac{R_{i}}{2\pi b}K_{0}\frac{r}{\lambda_{2}}
\end{equation} 
For small values of $r$,
\begin{equation}
\frac{V}{I_{o}}= \frac{R_{i}}{2\pi b}ln\frac{2 \lambda_{2}}{r}
\end{equation} 
For a three dimensional system [8 chapter 5 ]
\begin{equation}
 \frac{V}{I_{o}} = \frac{R_{i}}{4 \pi r} e^{\frac{-r}{\lambda}} = \frac{R_{i}}{4\pi r}( 1- \frac{r \sqrt{R_{i}\chi}}{\sqrt{R_{m}}} + ---)
\end{equation} 
As the dimensionality of the system increases, the input resistance becomes more and more insensitive to changes in $R_{m}$. In an infinite cable $ R_{in}\propto \sqrt {R_{m}}$, for a two dimensional sheet $ R_{in} \propto log (R_{m})$. In a three dimensional structure such as muscle tissue $R_{in} \propto e^({\frac{-1}{R_{m}^(1/2)}})$. 
\end{large}
\subsection{\textbf\it{$\lambda_{AC}$ : }}
\begin{large}
As $R_{m}$ and $C_{m}$ change with change in voltage in an active cable, $\lambda$ also changes. Thus, to get the actual value of $\lambda$ under these circumstances, one needs to look at the problem differently.The linear cable equation can be solved for measurements with sinusoidal currents. Here $R_{m}$ has to be replaced by $Z_{m}$(membrane impedance). The input impedance and length become frequency dependent. \\
The potential distribution in a linear cable when an a.c current is injected using intracellular electrodes can be written as [2]:
\begin{equation}
 V(x,t) = \frac{1}{2} (I_{0}(t)(Z_{m}r_{i})^{\frac{1}{2}})exp(\frac{-x}{(\frac{Z_{m}}{r_{i}})^{\frac{1}{2}} })
\end{equation} 
where $r_{i}= r_{a}$\\
In a.c analysis, $Z_{m}$ is replaced by a complex number so $\lambda_{AC}$ cannot have the same physical meaning as $\lambda_{DC}$ which is a real number. So a new $\lambda_{\omega}$ is defined.\\ 
Along a neurite, 
\begin{equation}
 \frac{V_{x}}{V_{o}} = e ^{-\frac{x}{\lambda \omega}} 
\end{equation} 
where \begin{equation}
       \lambda_{\omega} = \sqrt \frac{Z_{m}}{r_{i}}
      \end{equation} 
where $Z_{m}$ is the membrane impedance of a unit length of neurite and $r_{i}$ is the resistance of unit length of cytoplasm. If $f$ is large, transmembrane current is entirely capacitative. 
\begin{equation}
 Z_{m} \approx \frac{1}{j\omega C}
\end{equation} 
where $C$ is the capacitance of unit length of membrane 
\begin{equation}
 \frac{V_{x}}{V_{o}} \approx e^{x(1+j)\sqrt\frac{\omega Cr_{a}}{2}}
\end{equation} 
\begin{equation}
 \lambda_{\omega} \approx \frac{1}{\sqrt \frac{\omega Cr_{a}}{2}} \approx \sqrt \frac{2}{\omega Cr_{a}}
\end{equation} 
Substituting, $r_{a} = \frac{R_{i}}{\pi a^{2}}$ and $ C = 2\pi a C_{m}$ where $a$ is the radius of the neurite , $R_{i}$ is cytoplasmic resistivity in $\Omega cm $  and $C_{m}$ is specific membrane capacitance in $\mu Farad / cm^{2}$ in equation $27$
\begin{equation}
 \lambda _{\omega} \approx \sqrt \frac{2 \pi a^{2}}{\omega 2 \pi a C_{m}R_{i}} =  \sqrt \frac{a}{\omega C_{m}R_{i}} = \sqrt \frac{a}{2\pi fC_{m}R_{i}} = \frac{1}{2} \sqrt \frac{d}{\pi f R_{i}C_{m}}
\end{equation} 
where $d$ is the diameter of the dendrite.\\
Using this equation $22$ can be rewritten as :
\begin{equation}
 V_{pp} = \frac{1}{2} (I_{pp}(Z_{m}r_{i})^{\frac{1}{2}})exp(\frac{-x}{\lambda_{w}})
\end{equation} 
where $I_{pp}$ is the peak to peak current at the point of current injection. \\
Whenever fast changing membrane potentials are encountered either as action potentials or as injected pulses, $\lambda_{w}$ will determine the electrotonic spread due to the high frequency components of the fast changing voltages. \\
 Pettersen and Einevoll [13] express $\lambda_{AC}(\omega)$ for dendritic sticks of a finite length. 
\begin{equation}
 \lambda_{AC}(\omega) = \frac{\int_{0}^{l}z|\hat {i}_{m}(z)|dz}{\int_{0}^{l}|\hat{i}_{m}(z)|dz}
\end{equation} 
where $\omega = 2\pi f $ is the angular frequency, $|\hat{i}_{m}|$ is the amplitude of the sinusoidally oscillating current at position $ z$ when a sinusoidal current is injected at the soma. 
Equation $30$ reduces to the following for a dendrite of infinite length :
\begin{equation}
 \lambda _{AC}^{\infty}(\omega) = \lambda \sqrt(\frac{2}{1+\sqrt(1+(\omega \tau)^{2})})
\end{equation} 
where $\tau$ is the membrane time constant. $\lambda_{AC}$ is thus dependent on frequency. It decreases with increasing frequency. Pettersen and Einevoll[12] discuss in great length a low pass filtering effect of extracellular potentials as a result of this dependence of $\lambda_{w}$ on frequency.\\
It is seen that at high frequencies, transmembrane current is purely capacitative. At $ f  = \frac{1}{2}\pi\tau_{m} $, $R_{m}$ has no effect on propagation of signals $\geq 5f_{m}$. Here $\lambda_{\omega}$ is given by equation 28. This shows that $\lambda_{\omega}\propto \frac{1}{\sqrt f }$. This implies that at higher frequencies, the $\lambda_{\omega}$ is lower. 
\newline 
Using another approach,Lindsay and Rosenberg[10] show that an active neuron will have space and time constants which reflect dynamic biophysical properties of the membrane. The space constant $\Lambda$ and time constant $T$ give the spatial and temporal rates of exponential decay of the total membrane current during an action potential. To calculate this one needs to calculate conduction speed of the action potentials which can be done easily using extracellular methods. Thus one can get [10, equation 16] :
\begin{equation}
 \theta = \frac{1}{C_{m}}\sqrt{\frac{Ag_{a}\hat{g_{m}}}{P}}
\end{equation} 
where $\theta$ is the speed of movement of the action potential train in $cm/msec$, $P$ is the perimeter of the dendrite, $A$ is the crosssectional area, $C_{m}$ is the capacitance, $g_{a}$ is the axial conductance in $mS/cm$ and $\hat{g_{m}}$ is the membrane potential during an action potential. [10, equation 17] gives : 
\begin{equation}
 \Lambda = \sqrt{\frac{g_{a}d}{4\hat{g_{m}}}}
\end{equation} 
When $\hat{g_{m}} = g_{m}$, $\Lambda = \lambda_{DC}$
\end{large}
\vspace{2mm}
\newline
\subsection{\textbf\it{$\lambda_{ACtaper}$:}}
\subsubsection{\textbf\it{linear taper:}}
\begin{large}
\begin{displaymath}
 r(x) = \rho x +r_{0},
\rho = \frac{rl-r_{0}}{l} 
\end{displaymath}
where $r(x)$ is the radius at any given $x$ along the dendrite, $r_{0}$ is the radius of dendrite at point $0$ and $\rho$ is the linear taper. 
\begin{equation}
 \lambda_{ACtapered}= \frac{1}{2}(\frac{2r}{\pi fR_{i}C_{m}})^\frac{1}{2}
\end{equation} 
\begin{equation}
 (2)^2(\lambda_{ACtapered})^{2} = 2(\frac{r_{0} + \rho.x}{\pi fR_{i}C_{m}})
\end{equation} 
\begin{equation}
 \lambda_{ACtapered} = \frac{1}{2} (\frac{d_{0}}{\pi fR_{i}C_{m}})^{\frac{1}{2}}(1 + \frac{2\rho.x}{d_{0}})^{\frac{1}{2}}
\end{equation} 
\begin{equation}
 \lambda_{ACtapered}= \lambda_{AC}(1 + \frac{2\rho.x}{d_{0}})^{\frac{1}{2}}
\end{equation} 
\end{large}
\subsubsection{\textbf\it{exponential taper:}}
\begin{large}
 \begin{displaymath}
  r(x) = r_{0} exp(-\rho.x),
\rho = \frac{ln(r_{0}/rl)}{l}
 \end{displaymath}
where $r(x)$ is the radius at any given $x$ along the dendrite, $r_{0}$ is the radius of the dendrite at point $0$ and $\rho$ is the exponential taper.
\begin{equation}
 \lambda_{ACtapered}= \frac{1}{2}(\frac{2r}{\pi fR_{i}C_{m}})^\frac{1}{2}
\end{equation} 
\begin{equation}
 (2)^2(\lambda_{ACtapered})^{2}= 2\frac{r_{0}}{\pi fR_{i}C_{m}}(exp(-\rho.x))^{\frac{1}{2}}
\end{equation} 
\begin{equation}
  \lambda_{ACtapered} = \frac{1}{2} (\frac{d_{0}}{\pi fR_{i}C_{m}})^{\frac{1}{2}}(exp(-\rho.x))^{\frac{1}{2}}
\end{equation} 
\begin{equation}
 \lambda_{ACtapered} = \lambda_{AC}(exp(-\rho.x))^{\frac{1}{2}}
\end{equation} 
\end{large}
\section{\textbf{ Branched dendrites: }}
\subsection{\it{\textbf{$\lambda_{DC}$ Non tapering :}}}
\begin{large}
This problem can be solved with or without the use of equivalent cylinders. 
\end{large}
\subsubsection{\it{Without equivalent cylinders}:}
\begin{large}
Calculate $\lambda_{DC} $ for each section separately using the equation $5$. 
\end{large}
\vspace{2mm}
\newline
\subsubsection{\it{With equivalent cylinders}:}
\begin{large}
 Rall [22,24] showed that dendritic trees could be collapsed into a single equivalent cylinder provided they meet the following requirements [9]: \\
$1.$  $R_{m}$ and $R_{i}$ values are the same in all branches.\\ 
$2.$ All terminals have the same boundary condition.\\ 
$3.$ All terminal branches end at the same electrotonic distance $L$ from the origin in the main branch, where $L$ is the sum of the $L_{i}$ values from the origin to the distal end of every terminal. $L$ corresponds to the total electrotonic length of the equivalent cylinder.\\ 
$4.$ At every branch point, infinite input resistances must be matched. If all cables possess the same membrane resistance and intracellular resistivity, this implies
    \begin{equation}
     d_{o}^{3/2} = d_{1}^{3/2} + d_{2}^{3/2}
    \end{equation} 
where $d_{o}$ is the diameter of the parent branch and $d_{1}$ and $d_{2}$ are the diameters of the daughter branches. This last condition is called the $d^{3/2}$ law.\\ 
If these four conditions are met the equivalent cylinder can be considered to be a perfect representation of the entire tree provided current is injected in the initial terminal.
In case there is input to any of the daughter branches, an additional constraint should be obeyed :\\ 
$5.$ Identical synaptic inputs, whether current injection or conductance change, must be delivered to all corresponding dendritic locations. \\
In such a situation $ \lambda_{DC}$ can be calculated from equation $5$ where the values inserted should be that for the equivalent cylinder. 
The diameter $ D $ of the equivalent cylinder is given by :
\begin{equation}
D = \biggl[\sum_{j=1}^{n_{0}}d_{0j}^{3/2}\biggr]^{2/3}
\end{equation} 
$ D = d_{01}$ if there is only one dendrite emanating from the soma. 
\begin{equation}
 \lambda =  \biggl[(\frac{R_{m}}{4R_{i}})D\biggr]^{1/2}
\end{equation}
\end{large}
\vspace{2mm}
\newline
\subsection{\it{\textbf{$\lambda_{DC}$ - Tapering:}}}
\begin{large}
This problem too can be solved with or without the use of equivalent cylinders.
\end{large}
\vspace{2mm}
\newline
\subsubsection{\it{Without equivalent cylinders}:}
\begin{large}
$\lambda_{DC}$ is calculated for each individual branch using equation $7$. Here the values of $r$,$r_{o}$ and $\frac{dr}{dx}$ will vary from branch to branch. 
\end{large}
\vspace{2mm}
\newline
\subsubsection{\it{With equivalent cylinders}:}
\begin{large}
 Rall[22] has shown that under certain conditions, a tapering tree can be reduced to a one dimensional cylinder. The condition is [8, equation $ 7.43$]:
\begin{equation}
 nr^{3/2}\biggl[1 + \biggl[\frac{dr}{dx}\biggr]^{2}\biggr]^{1/4} = constant
\end{equation} 
where $n$ is the number of dendritic branches and $r$ the radius of all the branch segments. These are functions of actual distance $x$ from the soma. The following condition is to be met here [8,equation $7.45$]
\begin{equation}
 \frac{dA}{dx} \propto \frac{dZ}{dx}
\end{equation} 
where $A$ is the surface area of the dendrites. $Z$ is the electrotonic distance. 
For the dendritic tree equivalent tapering cable, the following conditions are to be met: 
\begin{equation}
 F(Z;K) = exp(K(Z-Z_{0}))
\end{equation} 
 \begin{equation}
  F(Z;K) = exp(KZ)
 \end{equation}
Then [14,equation $4$] gives : 
\begin{equation}
 n^{3/2}\biggl[1+ \biggl[\frac{dr}{dx}\biggr]^{2}\biggr]^{1/4} = (r_{0})^{3/2}n_{0} F(Z;K)
\end{equation}
where $n_{0}$ is the number of branches at $ x = 0$; $ F(Z;K)$ is the geometric ratio imposing a taper on the equivalent cable and $ K<0$ is the rate of taper.
\begin{equation}
 \frac{dA}{dx}\propto F(Z;K)\biggl[\frac{dZ}{dx}\biggr]
\end{equation}
If $\frac {dr}{dx} = 0$ and every branch at any given $x$ or $Z$ has a different diameter, then for $ F(Z;K)= exp(KZ)$ [22] 
Then [14,equation $6$] gives :
\begin{equation}
 F(Z;K) = \sum_{j=1}^{n(x)}(d_{j})^{3/2}\biggl[\sum_{j=1}^{n_{0}} (d_{j})^{3/2}\biggr]^{-1}
\end{equation}
$d_{j}$ is the diameter of the $jth$ branch at distance $x$ from soma. $ Z= 0$ when $ x = 0$.
Alternatively, if $x$ represents actual distance measured along successive branch points and branching occurs at distances $ 0 = x_{0}< ...< x_{p}$ with $ n_{i}$ branches between $x_{i}$ and $x_{i+1}$ where $ x_{i}\leq x\leq x_{i+1}$, then equation $51$ becomes [14,equation $7$]
\begin{equation}
 F(Z;K) = \sum_{j=1}^{n_{i}}d_{ij}^{3/2}\biggl[\sum_{j=1}^{n_{0}}d_{0j}^{3/2}\biggr]^{-1}, i = 0,1,...p
\end{equation}
 where $ (\frac{dr}{dx})^{2} \leq 1$ and all branches are equal in diameter [8 p- 156]. 
However, if $ F(Z;K) \neq 1$ \\ then [14 equation $ 9$] gives :
\begin{equation}
 D_{taper}= D \biggl[F(Z;K)\biggr]^{2/3 } = \biggl[\sum_{j = 1}^{n_{i}}d_{ij}^{3/2}\biggr]^{2/3}, i = 0,1,...p
\end{equation} 
From [14,equation $10$] we arrive at : 
\begin{equation}
 \lambda_{taper}= \biggl[\frac{R_{m}}{4R_{a}}D_{taper}\biggr]^{1/2}
\end{equation}
if there is profuse branching $ F(Z;K) >1 $ ,if there is paucity of branching $ F(Z;K)<1$
\newline
Jack et al[8] discusses the conditions underlying reducing a branched, tapering dendrite to an equivalent cylinder by taking three cases of taper and calculating the types of branching and values of Z that are possible with all three. To summarize here the types of taper considered are [8,equations $7.49,7.50,7.51$] :
\begin{equation}
 r = r_{0}(1-ax)
\end{equation} 
\begin{equation}
 r = r_{0}exp(-ax)
\end{equation} 
\begin{equation}
 r = r_{0}(1+ax)^{-1}
\end{equation} 
where $r_{0}$ is the initial radius and $a$ is the factor controlling the rate of taper with distance.If the dendrite is to be reduced to an equivalent cylinder the forms of branching need to be [8,equations $7.52,7.53,7.54$]:
\begin{equation}
 n = n_{0}(1-ax)^{-\frac{3}{2}}
\end{equation} 
\begin{equation}
 n = n_{0}e^{2ax}(\frac{a^{2}r_{0}^{2}+1}{a^{2}r_{0}^{2}+ e^{2ax}})^{\frac{1}{4}}
\end{equation} 
\begin{equation}
 n = n_{0}(1+ax)^{\frac{5}{2}}(\frac{a^{2}r_{0}^{2}+1}{a^{2}r_{0}^{2}+ (1+ax)^{4}})^{\frac{1}{4}}
\end{equation} 
where $n_{0}$ is the number of branches at $x=0$. \\
 The relationship between $Z$ and $x$ is an exact solution for taper described by equation ($55$). For the tapering conditions in equations ($56$) and ($57$) give approximate solutions [8,equations $7.55,7.57,7.58$]:
\begin{equation}
 Z= \frac{1}{\lambda_{0}}((1+a^{2}r_{0}^{2})^{\frac{1}{4}}\frac{2}{a}(1-(1-ax)^{\frac{1}{2}})
\end{equation} 
where 
\begin{equation}
  \lambda_{0} = (\frac{R_{m}r_{0}}{2R_{i}})^{\frac{1}{2}}
\end{equation} 
\begin{equation}
 Z \approx \frac{2}{a\lambda_{0}}(e^{(\frac{1}{2}ax)} -1)
\end{equation} 
\begin{equation}
 Z \approx \frac{2}{3a\lambda_{0}}((1+ax)^{\frac{3}{2}}-1)
\end{equation} 
\end{large}
\vspace{2mm}
\subsubsection{\it{Difference between $\lambda_{DC}$for full morphology and equivalent cylinder}:}
\begin{large}
 The equivalent cylinder formulae are strictly valid only for neurons that can be approximated as cylinders with uniform membrane resistivity. If the neuron under consideration has a dendritic taper  or a soma shunt, the formulas can lead to errors as there can be violations of the equivalent cylinder approximations. Even if these assumptions are satisifed, the accuracy of the parameters used in the formulae are also crucial [5]. Schierwagen[27] discusses the many situations in which neurons do not obey all the assumptions required before reduction to an equivalent cylinder. Spinal motoneurons and superior colliculus output neurons obey the $d^{\frac{3}{2}}$ rule but many others do not. One of the alternate models suggested is the morphology based branching cable model. It only expects that the condition requiring the same boundary condition at all branches be adhered to. \\
Given that most neurons do not follow the conditions in the equivalent cylinder approximations, the $\lambda_{DC}$ values calculated by the equivalent cylinder approach can at best be regarded as approximations. On the other hand, calculating the $\lambda_{DC}$ for the entire morphology of a complex neuron can be computationally expensive. Which approach is used may depend on weighing the errors versus efficiency needed in the particular problem. 
\end{large}

\subsection{\it{\textbf{$\lambda_{AC}$ Nontapering: }}}
\begin{large}
This problem can be solved without the use of equivalent cylinders and with the use of equivalent cylinders.
\end{large}
\vspace{2mm}
\newline
\subsubsection{\it{Without equivalent cylinders :}}
\begin{large}
Calculate $\lambda_{AC}$ branched for each branch using equation $28$ varying d, Ra and Cm where necessary.
\end{large}
\vspace{2mm}
\newline
\subsubsection{\it{With equivalent cylinders : }}
\begin{large}
Calculate $D$ from equation $43$. Then substitute that in equation $28$.
\begin{equation}
 \lambda_{ACbranchednontap} = \frac{1}{2}\sqrt{\frac{D}{\pi fR_{i}C_{m}}}
\end{equation} 
\end{large} 
\vspace{2mm}
\subsection{\it{\textbf{$\lambda_{AC}$ tapering: }}}
\begin{large}
This problem can be solved without the use of equivalent cylinders and with the use of equivalent cylinders.
\end{large}
\vspace{2mm}
\newline
\subsubsection{\it{Without equivalent cylinders :}}
\begin{large}
Calculate $\lambda_{ACtapered}$ branched for each branch using equations $37$ or $40$.
\end{large}
\subsubsection{\it{With equivalent cylinders : }}
\begin{large}
Calculate $Dtaper$ from equation $53$ and substitute it in equations $37$ or $40$.
\end{large}
\vspace{2mm}
\newline
\section{\textbf{Space constant - chemical:}}
\begin{large}
Calcium concentration in dendrites and spines is a function of diffusion, buffering and pumping. Under certain conditions this can be reduced to become similar to the linear one dimensional cable equation [30]. Thus quantities like space constant, time constant and input resistance can be defined for the reaction diffusion equation too. We are looking at the diffusion of calcium ions in a cylinder after the influx of the calcium current $ I_{Ca}(x,t)$ across the membrane. The radial components of diffusion are neglected so we get one dimensional flow. The calcium ions while diffusing bind to various buffers and can be pumped out of the cylinder. The buffer itself can diffuse with a diffusion coefficient $ D_{B}$. The following equations give the change in concentration of calcium and bound calcium buffer.
The one- dimensional diffusion equation is : 
\begin{equation}
 \frac{\partial C(x,t)}{\partial t} = D\frac{\partial^{2}C(x,t)}{\partial x^{2}} + \frac{2}{r} i(x,t)
\end{equation} 
where $C(x,t)$ is the concentration of $Ca^{2+}$ in $\mu M$ at time $t$ and position $x$ in response to the applied current density $i(x,t)$ $fA/\mu m^{2}$,$D$ is the diffusion constant in $\mu m^{2}/msec$ and the radius of the cable is $r$ $\mu m$.
A diffusable buffer is incorporated which shows second order kinetics :
\begin{equation}
 Ca^{2+} + B \xrightleftharpoons[b]{f} M 
\end{equation}
where $Ca^{2+}$ is the free calcium in $\mu M $, $B$ is the free buffer in $\mu M $ and $M$ is the bound buffer in $\mu M$. $f$ in $ msec^{-1}$ and $b$ in $\mu M msec^{-1}$ are the rate constants. 
Incorporating this into the diffusion equation we can get the following :
\begin{equation}
 \frac{\partial C}{\partial t} = D\frac{\partial^{2}C}{\partial x^{2}} - P(C) - fCB +bM +\frac{2}{r}i(x,t)
\end{equation} 
\begin{equation}
 \frac{\partial M}{\partial t} = D_{b}\frac{\partial^{2}M}{\partial x^{2}} + fCB -bM
\end{equation} 
\begin{equation}
 B_{T}= M(x,t) + B(x,t)
\end{equation} 
where $D_{b}$ is the diffusion constant of both the free and the bound buffer in $\mu m^{2}/msec$ and $B_{T}$ is the concentration of the total buffer. 
Under certain limiting conditions of low $[Ca^{2+}]$ and fast kinetics of buffering compared to kinetics of diffusion, the above sets of nonlinear equations can be reduced to a single linear equation similar to the cable equation [30]. When an externally applied point source current term is used $I(x,t)$, the resulting equation is [30, equation 13] :
\begin{equation}
 r(\frac{1+\beta}{2})\frac{\partial C}{\partial t}= r(\frac{D+\beta D_{b}}{2})\frac{\partial^{2}C}{\partial x^{2}} -P_{m}C(x,t)+K_{\infty}P_{m}I(x,t)
\end{equation} 
where $\beta =\frac{ B_{T}}{K_{d}}$, $P_{m}$ is the membrane pump parameter at $\mu m/msec$ and $K_{\infty}$ is the constant of proportionality. This can be compared to the cable equation :
\begin{equation}
 C_{m}\frac{\partial V(x,t)}{\partial t}= \frac{r}{2R_{i}}\frac{\partial^{2}V}{\partial x^{2}} - \frac{1}{R_{m}}V(x,t) + \frac{R_{\infty}}{R_{m}}I(x,t)
\end{equation} 
where $V(x,t)$ is the voltage, $C_{m}$ is the capacitance, $R_{i}$ is the axial resistance, $R_{m}$ is the membrane resistance, $R_{\infty}$ is the input resistance.
Both equations are similar as can be seen by [30, equation 15] :
\begin{equation}
R_{m}^{-1}  \leftrightarrow P_{m} ; C_{m} \leftrightarrow r\frac{(1+\beta)}{2}; R_{i}^{-1} \leftrightarrow D +\beta D_{b}; R_{\infty} \leftrightarrow K_{\infty}                                         
\end{equation} 
Under this condition the space constant can be defined as [30,equation 16]:
\begin{equation}
 \lambda_{C}= \sqrt \frac{r(D+\beta D_{b})}{2P_{m}}
\end{equation} 
Iannela and Tanaka[6] discuss in detail the general form of a nonlinear cable equation with calcium which is then reduced to a linear cable equation. The sodium calcium exchanger is excluded and the voltage dependent calcium channels are assumed to be independent of calcium. This along with the rapid buffer approximation where $[Ca]_{i}\ll K_{d}$, gives rise to a pump which extrudes endogeneous calcium in a linear fashion. The buffer is also assumed to reach equilibrium much faster than the diffusion of calcium. This leads to a closed form relation between the rates of change of $[Ca]_{i}]$ and $M$. Thus the reaction diffusion equation can be reduced to the chemical cable equation ($63$). 
\section{\textbf{Discussion}}
In this paper an attempt has been made to describe lambda ( also called space constant/ variable space constant/ length parameter) under different conditions. Intuitively, the notion of a space constant or a variable space constant is more obvious in the case of a passive dendrite and often the equation to estimate this for passive cases is used for active cases too [26]. As described earlier, the space constant in a cable ( infinite, semi infinite and finite) gives an indication of the extent of voltage decay along the cable. The larger the diameter, the greater the passive spread of voltage. It also influences the summation of synaptic inputs. Spatially separated synaptic inputs in smaller diameter cables will sum differently than those in larger diameter cables. Once again it is important to distinguish between morphological length and the electrical length or electrotonic length of the dendrite. Thus in a given neuron, there could be different tree sizes but they could be of similar electrotonic lengths due to the differences in the diameter and the resulting space constant. As discussed by [28,(Chapter-5)],`` The depolarization at the soma in response to a given input current occuring at any point on any dendritic tree, depends, for a given neuron, only on the electrotonic distance between input and soma. The response is independent of which dendritic tree receives the input and is independent of the geometrical details of the tree that receives the input and the geometrical details of all the other dendritic trees. The magnitude of the response is inversely proportional to the sum of the conductances per characteristic length(space constant) of the dendritic trunks.''  \\
In the case of active conduction, the greater the frequency, the smaller is the $\lambda$ and thus greater is the attenuation of voltage. Here $\lambda$ is a function of $V$ and this introduces nonlinearity into the system. However, by making certain approximations as shown by [15,17,6,7] the nonlinearities can be reduced to linearities by using the ionic cable theory. This involves postulating the nonlinear component at discrete locations and mathematically introducing a Dirac delta function in the cable equation. With this transformation, length parameter used for the passive case can still be used in the active case. \\
Reiterating the relationship between $\lambda$ and input resistance it can be said that $\lambda$ is directly proportional to input resistance in a finite and semi - infinite cable. It is proportional to $\sqrt Z$ ( impedance) in an active cell or where current injection is sinusoidal. \\
Goldstein and Rall[3] have shown that diameter,tapering and branching are important factors in determining action potential propagation. They defined a geometric ratio (GR) which is $\Sigma d_{j} ^{\frac{3}{2}} / d_{a}^{\frac{3}{2}}$ where $d_{a}$ is the diameter of the parent branch and $d_{j}$ is the diameter of the daughter branch. For uniform cables, if $GR=1$, there is an impedance match and propagation is not affected. If $GR<1$, a favorable impedance mismatch occurs and action potentials propagate effectively. However if $GR>1$, the impedance mismatch is unfavorable and action potential propagation is inefficient. Vetter et al[29] showed in a simulation study which isolated morphology as the only variable, that backpropagation of action potentials was correlated with membrane area distribution in the dendritic tree and the GR at individual branch points. The same study showed that in cells with elaborate dendritic trees like Purkinje cells, backpropagation is insensitive to the sodium channel density over the physiological range. However in dopamine neurons, even low sodium channel density leads to efficient backpropagation. They also demonstrate that dendritic geometry places a limit on modulation of backpropagation by channel density and neurotransmitters. \\
Bernander et al[1] and  Rapp et al[25], show in two set of simulation studies that any individual cell is dependent on the network activity in which it is embedded. The former study simulated a layer $5$ cortical pyramidal cell receiving inputs from 4000 excitatory and 1000 inhibitory cells firing spontaneously between $0-7$ Hz. Here $\tau_{m}$ and $R_{in}$ change by a factor of $10$ ( $80-7$msec and $110-14$ Mohms) while the electrotonic length of the cell changes by a factor of $3$. In the [25] study,which modeled a Purkinje cell with parallel fibres, it was also seen that even at a low firing rate of a few Hz, the parallel fibre activity changed the membrane conductance of the Purkinje cell. The time constant $\tau_{m}$ and the input resistance $R_{in}$ decrease several fold while electrotonic length $L$ and the voltage attenuation factor increase significantly. This in turn affects the spatial and temporal processing of individual neurons. Thus results from a slice preparation and that from \textit{in vivo} recording can give us different answers about the functions of an individual neuron. In this light, the electrotonic properties of the cell are not static but dynamic, evolving with the background environment of the cell. \\ 
In summary, it could be said that random synaptic inputs on a dendrite will cause a depolarization at the soma depending on the electrotonic distance in the passive case. This in turn is dependent on the space constant of the dendrite. In the active case, the propagation of action potentials depends on the membrane area ratio and GR in the dendritic tree. This in turn is dependent on the ratio of diameters of parent and daughter branches. In other words, geometry does play a role here. In the active case there is an inverse relation between $\lambda_{\omega}$ and the frequency of the signal. Finally, it is important to take into account the role played by random, background synaptic activity on the neuron. As the synaptic activity is enhanced ( either by increase in numbers of synapses, or frequency), the electrotonic length and effective membrane time constant are both increased aiding in synaptic integration. 
\end{large}
\section*{Acknowledgments}
AG will like to acknowledge the comments made by Prof.R. Poznanski on an earlier draft. AG also acknowledges Prof.A.K Gupta,former Head, Dept of Imaging Sciences and Interventional Radiology, SCTIMST currently at NIMHANS, Bengaluru for providing her a base to work in SCTIMST. This work has been done with funds from the WOS- A grant of the Department of Science and Technology, India. This is a scheme meant to support women scientists.
\vspace{2mm}
%\section*{\begin{large}\textbf{References :}\end{large}}

\vspace{2mm}
\end{document}